\documentclass[%
 reprint,
 amsmath,amssymb,
 aps,
]{revtex4-1}

\usepackage{graphicx}
\usepackage{dcolumn}
\usepackage{bm}

\usepackage{graphicx}
\usepackage{amsmath}
\usepackage{amssymb}
\usepackage{amsfonts}
\usepackage{hyperref}
\usepackage{url}
\usepackage{color}
\usepackage{bm}
\usepackage{array}
\usepackage{tabularx}

\newcommand{\apjl}{ApJ}

\newcommand{\physrep}{Physics Report}
\newcommand{\jcap}{JCAP}

\newcommand{\be}{\begin{equation}}
\newcommand{\ee}{\end{equation}}
\newcommand{\ba}{\begin{eqnarray}}
\newcommand{\ea}{\end{eqnarray}}
\newcommand{\nn}{\nonumber}

\newcommand{\mpch}{{\mathrm{Mpc}}\,h^{-1}}
\newcommand{\hmpc}{h\,{\mathrm{Mpc}}^{-1}}

\begin{document}

\preprint{APS/123-QED}

\title{\boldmath Non-Linearity-Free prediction of the growth-rate $f\sigma_8$ \\ using Convolutional Neural Networks}

\author{Koya Murakami$^1$}
\email{koya.murakami9627@gmail.com}
\author{Indira Ocampo$^2$}
\email{indira.ocampo@csic.es}
\author{Savvas Nesseris$^2$}
\email{savvas.nesseris@csic.es}
\author{Atsushi J. Nishizawa$^{3,4,5}$}
\email{atsushi.nisizawa@gifu.shotoku.ac.jp}
\author{Sachiko Kuroyanagi$^{2,1}$}
\email{sachiko.kuroyanagi@csic.es}

\affiliation{
$^1$Department of Physics, Nagoya University, Furocho, Chikusa, Nagoya, 464-8602, Japan\\
$^2$Instituto de F\'isica Te\'orica UAM-CSIC, Universidad Auton\'oma de Madrid,
Cantoblanco, 28049 Madrid, Spain\\
$^3$Gifu Shotoku Gakuen University, Takakuwa-Nishi, Yanaizu, Gifu, 501-6194, Japan\\
$^4$Institute for Advanced Research, Nagoya University, Furocho, Chikusa, Nagoya, Aichi, 464-8602, Japan\\
$^5$Kobayashi Maskawa Institute, Nagoya University, Furocho, Chikusa, Nagoya, Aichi, 464-8602, Japan
}

\date{\today}

\begin{abstract}
The growth-rate $f\sigma_8(z)$ of the large-scale structure of the Universe is an important dynamic probe of gravity that can be used to test for deviations from General Relativity. However, for galaxy surveys to extract this key quantity from cosmological observations, two important assumptions have to be made: i) a fiducial cosmological model, typically taken to be the cosmological constant and cold dark matter ($\Lambda$CDM) model and ii) the modeling of the observed power spectrum, especially at non-linear scales, which is particularly dangerous as most models used in the literature are phenomenological at best. In this work, we propose a novel approach involving convolutional neural networks (CNNs), trained on the Quijote N-body simulations, to predict $f\sigma_8(z)$ directly and without assuming a model for the non-linear part of the power spectrum, thus avoiding the second of the assumptions above. This could serve as an initial step towards the future development of a method for parameter inference in Stage IV surveys. We find that the predictions for the value of $f\sigma_8$ from the CNN are in excellent agreement with the fiducial values since they outperform a maximum likelihood analysis and the CNN trained on the power spectrum.
Therefore, we find that the CNN reconstructions provide a viable alternative to avoid the theoretical modeling of the non-linearities at small scales when extracting the growth rate.
\end{abstract}

\maketitle


\section{Introduction}
Recent cosmological observations indicate that the Universe is experiencing accelerated expansion, typically attributed to a new form of matter, i.e., \textit{dark energy}, that is responsible for this phenomenon. Within the context of the Friedmann-Lema\^itre-Robertson-Walker (FLRW) cosmology, this new dark energy component is frequently attributed to the cosmological constant $\Lambda$, which is introduced regardless of the fine-tuning issues that this may cause~\cite{padmanabhan2003cosmological, amendola2013internal}. Together with the assumption of a cold dark matter (CDM) component, these two ingredients comprise the so-called $\Lambda$CDM model, which is currently the concordance cosmological model~\cite{Planck:2018vyg}.

The large-scale structure (LSS) of the Universe encodes very important details crucial to testing cosmological models, as it provides information about the late-time evolution of the Universe and the structure of the matter density field. Both are useful for constraining the current values of the fractional density parameters for CDM, baryonic matter, and dark energy: $\Omega_\mathrm{m,0}$, $\Omega_\mathrm{b,0}$ and $\Omega_\mathrm{\Lambda,0}$ respectively, and many other parameters like the clustering strength $\sigma_{8}$~\cite{DES:2021wwk,kacprzak2022deeplss}. They are also important probes to search for possible deviations from general relativity.

An overriding challenge in modern cosmology is understanding the features of dark energy, and here lies the importance of obtaining highly accurate data from forthcoming cosmological missions such as \textit{Euclid}~\cite{EUCLID:2011zbd}, LSST~\cite{LSSTScience:2009jmu}, and DESI~\cite{DESI:2016fyo}, which aim to provide measurements to a few percent of the various key parameters related to the LSS. In order to make sense of the plethora of currently available cosmological models and observations~\cite{Joyce:2016vqv, Motta:2021hvl}, it is necessary to develop valuable statistical tools. In particular, Machine Learning (ML) has attracted increasing attention as it simplifies the usually computationally expensive procedures for data treatment~\cite{lazanu2021extracting}. 
Simulations also play an indispensable role in understanding the whole picture of the LSS since some information is particularly difficult to extract due to the non-linear nature of the system. This is the motivation behind the Quijote N-body simulations, consisting of CDM particle simulations that allow quantifying data on the matter field in the fully non-linear regime and with different statistics ~\cite{villaescusa2020quijote}.

One way of probing the dynamics of LSS is by measuring the growth of matter density perturbations $\delta_{m}=\delta\rho_{m}/\rho_m$, where $\rho_m$ is the background matter density and $\delta\rho_{m}$ its linear order perturbation, and its growth rate that is usually represented by its logarithmic derivative $f=\frac{d \ln \delta_m}{d \ln a}$. However, when dealing with galaxy surveys, the observed quantity is, in fact, the galaxy density $\delta_g$, which is connected to the matter perturbations through a bias parameter that changes from survey to survey: $\delta_{g}=b\,\delta_{m}$ and depending on the type of galaxy observed~\cite{Desjacques:2016bnm}. At some redshift bin $z_i$, the growth rate $f(z_i)$ can be combined with the bias $b(z_i)$ to give rise to the so-called velocity-density coupling parameter $\beta(z_i)=f(z_i)/b(z_i)$. At the same time, the combination of $b(z_i)\sigma_8(z_i)$ can also be independently measured~\cite{sagredo2018internal}, where $\sigma_8(z)$ is the root mean square (RMS) density fluctuation in a sphere of radius $R=8\ \mpch$. Therefore, the combination $f\sigma_8(z) \equiv f(z)\sigma(z)$ is independent of bias and can be measured via redshift-space distortions~\cite{song2009reconstructing}. Additionally, other approaches were developed in recent years: analyses of galaxy clustering have also been done using the Effective Field Theory of Large-Scale Structure (EFTofLSS)~\cite{Carrasco:2013mua, Carrilho:2022mon} and simulation-based emulators were implemented ~\cite{Winther:2019mus, Ramachandra:2020lue, Brando:2022gvg}.

In order to extract $f\sigma_8(z)$ with this approach, two main assumptions have to be made: first, that of a cosmological model (typically assumed to be the $\Lambda$CDM model), as we need to convert the redshifts of galaxies and coordinates in the sky to distances, for extracting the correlation function~\cite{WiggleZ:2013akc}. Second, we need to assume a model for the non-linear part of the power spectrum, typically done with phenomenological models~\cite{Scoccimarro:1995if, Bernardeau:2011vy, Baldauf:2015xfa, Casas:2017eob, Euclid:2019clj}.

This approach has also created several $f\sigma_8(z)$ dataset compilations; see, for example, Refs.~\cite{sagredo2018internal, Nesseris:2017vor, Kazantzidis:2018jtb} and references therein, where the data explicitly depend on the cosmology used. This can be somewhat easily corrected via an Alcock-Paczynski (AP) type correction~\cite{Nesseris:2017vor}, but the second issue of the model dependence on the non-linear power spectrum is more insidious as all models used are phenomenological, and thus far, there is no way to correct for that a posteriori.

One approach that has been recently proposed to extract cosmological parameters in a theory-agnostic manner is via ML techniques, in particular, by training Convolutional Neural Networks (CNNs) on N-body simulations and then extracting the parameters from the LSS statistics~\cite{Lazanu:2021tdl}. This approach was also further expanded to extract the cosmological density and velocity fields from N-body simulations~\cite{Wu:2023wmj,Qin:2023dew}. These ML methods have the main advantage that after the original training has occurred, any subsequent evaluations are practically instantaneous, and more importantly, they allow for extracting the quantities of interest without assuming a specific functional form for the power spectrum.

Therefore, in this work, we aim to extract and compare the growth-rate $f\sigma_8$ by leveraging ML techniques. We implemented and trained a CNN directly on N-body simulations at different redshift bins and also, on the power spectrum of these simulations. 
While we do not steer clear of the cosmological model premise, we refrain from assuming the modelling of the non-linear power spectrum.

Lastly, let us clarify that, even if one could measure a dark matter (DM) density map, the analysis would not be as direct and useful as the one with $f\sigma_8$ measurements in the end. The reason is that while the DM density map serves as a valuable tool, model comparison, e.g. $\Lambda$CDM versus $f(R)$, is typically conducted using $f\sigma_8$, which essentially compresses the same information. Therefore, having predictions for $f\sigma_8$ in a manner as model-independent as possible is highly desirable. An approximate analogy is found in the Planck CMB maps at the pixel level. While these maps are useful for analysis and contain comprehensive cosmological data, parameter inference and model selection are primarily based on the angular power spectra.

The layout of our paper is as follows: in Sec.~\ref{sec:theory}, we briefly summarize the theoretical background of our analysis, while in Sec.~\ref{sec:quijote}, we present the details for the Quijote simulations used in our analysis. Then, in Sec.~\ref{sec:density}, we describe the results of our ML analysis using the density field of the simulations. Finally, we summarise our conclusions in Sec.~\ref{sec:conclusions}.

\section{Theoretical framework \label{sec:theory}}
\subsection{The $\Lambda$CDM model and the growth-rate}
In this work, we will consider general relativity and a flat $\Lambda$CDM universe with an equation of state of $w=-1$ for dark energy. 

At the background level, the Hubble parameter in a flat $\Lambda$CDM model is given by the first Friedmann equation as usual:
\begin{equation}
H(a)^2 = H_0^2\,\left( \Omega_{m,0}\,a^{-3}+1-\Omega_{m,0} \right),\label{eq:hubble}
\end{equation}
where $H_0$ is the Hubble constant and the matter density $\Omega_{m,0}$ can be related to the scale factor $a$ by:
\begin{equation}
\Omega_{m}(a) = \frac{\Omega_{m,0}\,a^{-3}}{{H(a)^2}/{H_0^2}}.
\end{equation}
Also, assuming a flat universe, we can calculate the comoving distance $D(z)$ from us to a redshift $z$ as
\begin{equation}
    D(z) = \int^z_0 \frac{c\ dz'}{H(z')},
\end{equation}
where $c$ is the speed of light and $H(z)$ is Hubble parameter at redshift $z$ calculated via Eq.~\eqref{eq:hubble}, when neglecting radiation and neutrinos at late times.

On the other hand, observations from the LSS and the cosmic microwave background (CMB) also suggest the existence of small $\sim \mathcal{O}(10^{-5})$ perturbations. Consequently, we have to work in the framework of a perturbed FLRW metric, as the gravitational instability produced by density fluctuations plays a crucial role in seeding the structures at large scales. In the conformal Newtonian gauge, we consider scalar metric perturbations $\Psi$ and $\Phi$, so the perturbed metric can be written as~\cite{ma1995cosmological,bernardeau2002large}
\begin{equation}
d s^2=a(\tau)^2\left\{-[1+2\,\Psi(\vec{x}, \tau)]\,d \tau^2+[1-2\,\Phi(\vec{x}, \tau)]\,d \vec{x}^2\right\},
\end{equation}
where the potentials depend on the space-time point $x^\mu=(\vec{x}, \tau) $, with $\tau$ being the conformal time, while the scale factor $a(\tau)$ only depends on the conformal time.

In general, we can assume that the matter component behaves as a perfect fluid and is described by the stress-energy tensor:
\begin{equation}
T_\nu^\mu=P \delta_\nu^\mu+(\rho+P) U^\mu U_\nu,
\end{equation}
where the 4-velocity is $U^\mu=\frac{d x^\mu}{\sqrt{-d s^2}}$, the total density is $\rho = \bar{\rho}+\delta \rho$, the total pressure is $P = \bar{P}+\delta P$, also $\delta \rho = \delta \rho(\vec{x},\tau)$ and $\delta P = \delta P (\vec{x},\tau)$ are the density and pressure perturbations respectively, while $\bar{\rho}=\bar{\rho}(\tau)$ and $\bar{P}=\bar{P}(\tau)$ are the background energy density and pressure quantities. Therefore, the stress-energy tensor components are~\cite{nesseris2022effective}:
\ba
T_0^0 &=&-(\bar{\rho}+\delta \rho), \\
T_i^0 &=&(\bar{\rho}+\bar{P}) u_i, \\
T_j^i &=&(\bar{P}+\delta P) \delta_j^i+\Sigma_j^i,
\ea
where $\Sigma_j^i \equiv T_j^i-\delta_j^i\,T_k^k / 3$ is the anisotropic stress and $u=\dot{\vec{x}}$. The dot denotes the derivative with respect to $\tau$~\cite{kurki2005cosmological}. Recall that the energy-momentum tensor follows the conservation law $ T^{\mu \nu}{ }_{; \nu}=0$, as a consequence of the Bianchi identities~\cite{nesseris2022effective}.

To study the evolution of the perturbed variables, we resort to the perturbed Einstein equations in $k$-space~\cite{ma1995cosmological, nesseris2022effective}:
\begin{equation}
k^2 \Phi+3 \frac{\dot{a}}{a}\left(\dot{\Phi}+\frac{\dot{a}}{a} \Psi\right)=4 \pi G_{\mathrm{N}} a^2 \delta T_0^0, 
\end{equation}
\begin{equation}
k^2\left(\dot{\Phi}+\frac{\dot{a}}{a} \Psi\right)=4 \pi G_{\mathrm{N}} a^2(\bar{\rho}+\bar{P}) \theta,
\end{equation}
\begin{align}
\ddot{\Phi}+\frac{\dot{a}}{a}(\dot{\Psi} +2 \dot{\Phi})+\left(2 \frac{\ddot{a}}{a}-\frac{\dot{a}^2}{a^2}\right) \Psi &+\frac{k^2}{3}(\Phi-\Psi) \nn \\ &=\frac{4 \pi}{3} G_{\mathrm{N}} a^2 \delta T_i^i,
\end{align}
\begin{equation}
k^2(\Phi-\Psi)=12 \pi G_{\mathrm{N}} a^2(\bar{\rho}+\bar{P}) \sigma,
\end{equation}
where the fluid velocity is defined via $\theta = ik_j u^j$ and $k_j$ is the wavenumber of the perturbations in Fourier space. We can also rewrite the anisotropic stress as $(\bar{\rho}+\bar{P}) \sigma \equiv - ( -\hat{k}_i\hat{k}_j-\frac{1}{3}\delta_{ij}) \Sigma^{ij}$. 

By taking the following approximations: sub-horizon (only the modes in the Hubble radius are important) and quasi-static (neglect terms with time derivatives), we can simplify the perturbed Einstein equations. Let us consider the perturbation of the Ricci scalar and see how it simplifies with these approximations:
\begin{equation}
\begin{aligned}
\delta R & =-\frac{12\left(\mathcal{H}^{2}+\dot{\mathcal{H}}\right)}{a^{2}} \Psi-\frac{4 k^{2}}{a^{2}} \Phi+\frac{2 k^{2}}{a^{2}} \Psi \\
& \quad -\frac{18 \mathcal{H}}{a^{2}} \dot{\Phi}-\frac{6 \mathcal{H}}{a^{2}} \dot{\Psi}-\frac{6 \ddot{\Phi}}{a^{2}},\\
\delta R & \simeq-\frac{4 k^{2}}{a^{2}} \Phi+\frac{2 k^{2}}{a^{2}} \Psi,
\end{aligned}
\end{equation}
then, we can find the following expressions for the Newtonian potentials, see~\cite{nesseris2022effective}:\begin{gather}
\Psi(k,a)=-4 \pi G_{\mathrm{N}} \frac{a^{2}}{k^{2}} \mu(k,a) \bar{\rho}_{\mathrm{m}} \delta_{\mathrm{m}}, \\
\Phi(k,a)=-4 \pi G_{\mathrm{N}} \frac{a^{2}}{k^{2}} Q_{\mathrm{eff}}(k,a) \bar{\rho}_{\mathrm{m}} \delta_{\mathrm{m}},
\end{gather}
where $\mu(k,a)\equiv G_\mathrm{eff}(k,a)/G_\mathrm{N}$ is used to denote an evolving Newton's constant. In GR, the two parameters $\mu(k,a)$ and $Q_{\mathrm{eff}}(k,a)$ can be shown to be equal to unity, but in modified gravity theories they are, in general, time and scale-dependent~\cite{nesseris2022effective}.

From the context developed before and by using the continuity equations that come from the conservation of the energy-momentum tensor (via the Bianchi identities), we arrive at a second-order differential equation that describes the evolution of the  matter density perturbations (in the absence of massive neutrinos), which is valid in the context of most modified gravity models~\cite{tsujikawa2007matter}:
\begin{align}
\delta^{\prime\prime}_\mathrm{m}(a)+&\left( \frac{3}{a}+\frac{H^{\prime}(a)}{H(a)} \right) \delta^{\prime}_\mathrm{m}(a) \nn \\ 
&-\frac{3\Omega_\mathrm{m,0}\,\mu(k,a)}{2 a^5\,H(a)^2/H_0^2} \delta_\mathrm{m}(a)=0,\label{eq:growth-ode}
\end{align}
for which we assume the initial conditions $\delta_\mathrm{m}(a\ll1)\sim a$ and $\delta'_\mathrm{m}(a\ll1)\sim 1$ at some initial time in the matter domination era, e.g. $a=10^{-3}$.

In GR and the $\Lambda$CDM model ($\mu=1$), the analytical solution for the growing mode can be found by directly solving Eq.~\eqref{eq:growth-ode} and is given by~\cite{analytic_fsigma8,Silveira:1994yq,Percival:2005vm,Buenobelloso:2011sja}
\begin{equation}
\label{eq:delta_m}
\delta_\mathrm{m}(a)=a \cdot{ }_2 F_1\left(\frac{1}{3}, 1 ; \frac{11}{6} ; a^3\left(1-\frac{1}{\Omega_\mathrm{m, 0}}\right)\right),
\end{equation}
where we use the quasi-static and sub-horizon approximations~\cite{tsujikawa2007matter} and ${}_2 F_1$ is the Gauss hypergeometric function expressed as\footnote{In the \texttt{scipy} python module, the Gauss hypergeometric function ${}_2 F_1$ is implemented as scipy.special.hyp2fi(a,b,c,z).}
\begin{equation}
    {}_2 F_1(a,b,c,z) = \sum^{\infty}_{n=0} \frac{(a)_n (b)_n}{(c)_n} \frac{z^n}{n!},
\end{equation}
where $(x)_n$ is the rising factorial calculated as
\begin{equation}
    (x)_n = x (x+1) (x+2) \cdots (x+n-1).
\end{equation}

With this in mind, we can define several key quantities: the growth rate $f$, and $\sigma_8$, which is the root mean square (RMS) normalization of the matter power spectrum as:
\begin{align}
f(a) & =\frac{\mathrm{d} \log \delta_\mathrm{m}}{\mathrm{~d} \log a}, \\
\sigma_8(a) & =\sigma_{8,0} \frac{\delta_\mathrm{m}(a)}{\delta_\mathrm{m}(1)}, \\
\sigma_{8,0}^2 & =\left\langle\delta_\mathrm{m}(x)^2\right\rangle.
\end{align}

It's important to mention that, in general, we have to correct the data for the Alcock-Paczynski effect (as different surveys use different fiducial cosmologies)~\cite{li2019redshift, li2018cosmological,Kazantzidis:2018rnb}. The combination of the growth rate $f(z)$ and $\sigma_8(z)$ gives rise to a new bias-independent variable $f \sigma_8$, that is, in fact, what is measured by galaxy surveys from redshift space distortions (RSDs): 
\begin{equation}
f \sigma_8(a)=a \frac{\delta_\mathrm{m}^{\prime}(a)}{\delta_\mathrm{m}(1)} \cdot \sigma_{8,0}.
\end{equation}
This quantity can be directly measured from current and forthcoming galaxy surveys, and several compilations exist in the literature~\cite{sagredo2018internal,Nesseris:2017vor,Kazantzidis:2018jtb}. Thus, if a growth rate measurement has been obtained via a fiducial cosmology $H'(z)$, then the corresponding $f\sigma_8$ value is obtained for the true cosmology
$H(z)$ via  an Alcock-Paczynski-like correction~\cite{Macaulay:2013swa,Nesseris:2017vor, Kazantzidis:2018rnb}:
\be 
\label{eq:calc_fs8}
f\sigma_8(z)\simeq \frac{H(z) D_A(z)}{H'(z) D_A'(z)} \,f\sigma_8'(z).
\ee 

Nonetheless, one of the main advantages of the $f \sigma_8$ growth rate is that it is a direct dynamic probe of gravity since, as can be seen from Eq.~\eqref{eq:growth-ode}, the dependence on the gravitational theory appears explicitly via the normalized evolving Newton's constant $\mu(k,a)$ and indirectly via the Hubble parameter $H(a)$. On the other hand, one of its main weaknesses is also that the measurements, as currently made by the galaxy surveys, suffer from model dependence (typically assumed to be the $\Lambda$CDM model) and by the fact that the modelling of the non-linear scales is phenomenological~\cite{Scoccimarro:1995if,Bernardeau:2011vy,Baldauf:2015xfa,Casas:2017eob,Euclid:2019clj}.

\subsection{Modeling of the non-linear scales}

While this work aims to predict $f \sigma_8$ without assuming a specific model, for comparison with the machine learning approaches, we performed a maximum likelihood analysis for the Effective Field Theory (EFT) power spectrum in redshift space in a later section. Here, we provide a brief review of this approach for completeness.

The power spectrum analysis needs to adopt a theoretical model to accurately account for the non-linear effects in galaxy clustering~(\cite[e.g.][]{Tegmark:1997rp, Seo:2003pu,Yahia-Cherif:2020knp,Euclid:2019clj, 2021JCAP...03..100C, 2021JCAP...04..039A, 2021JCAP...11..028A, 2019JCAP...03..007V, 2016arXiv161009321P, 2010PhRvD..82f3522T, 2004PhRvD..70h3007S, 1984ApJ...284L...9K}). We introduce the non-linear power spectrum, incorporating 1-loop corrections, through the utilization of the EFT power spectrum in redshift space, described as~\citep{2022JCAP...11..038N}
\begin{align}
\label{eq:EFT_pk}
    P^{\rm EFT}_s (k, \mu) &= P_{\delta \delta} (k) + 2 f_0 \mu^2 P_{\delta \theta} (k) + f^2_0 \mu^4 P_{\theta \theta} (k) \nonumber \\
    &+ A^{\rm TNS} (k, \mu) + D(k, \mu) \nonumber \\
    &+ (\alpha_0 + \alpha_2 \mu^2 + \alpha_4 \mu^4 + \alpha_6 \mu^6) k^2 P_L (k) \nonumber \\
    &+ \tilde{c}(f_0 \sigma_v k \mu)^4 P_s^{K} (k, \mu) \nonumber \\
    &+ P_{\rm shot} [ \alpha^{\rm shot}_{0} + \alpha^{\rm shot}_2 (k \mu)^2 ] ,
\end{align}
where $\mu$ is the cosine of the angle between the line of sight, and $f_0$ is the growth rate at large scales. 

The first and second lines of the equation represent the power spectrum in redshift space up to the 1-loop order in perturbation theory. Here, $P_{\delta \delta}$ and $P_{\theta \theta}$ refer to the 1-loop power spectra for the density and velocity fields in real space, respectively. The term $P_{\delta \theta}$ denotes the cross-power spectrum between these fields. Additionally, $A^{\rm TNS} (k, \mu)$ and $D(k, \mu)$ are correction terms derived from combinations of three and four fields (densities or velocities), respectively~\citep{2010PhRvD..82f3522T, 2021JCAP...11..028A}.

In the third line, $P_L (k)$ represents the linear matter power spectrum, and the EFT parameters $\alpha_0$, $\alpha_2$, $\alpha_4$, and $\alpha_6$ are introduced. This term accounts for small-scale variations in the behaviour of large scales and the mapping between real and redshift space in the non-linear regime. The forth line represents the Finger-of-God effects \citep{2020JCAP...05..042I}, where $\sigma_v$ is the velocity dispersion and $P^K_s (k, \mu)$ is the Kaiser power spectrum \citep{1984ApJ...284L...9K}.

The last line accounts for noise effects. Here, $P_{\rm shot}$ denotes the Poisson shot noise, which is proportional to $1/n$, where $n$ represents the number density of the tracer. The term $(k \mu)^2$ characterizes deviations from pure white noise  \citep{2009JCAP...08..020M, Desjacques:2016bnm, 2020JCAP...07..062C}.

\section{The Quijote simulations \label{sec:quijote}}
The Quijote simulations, see Ref.~\cite{villaescusa2020quijote}, are a set of 44100 N-body simulations created via the TreePM code Gadget-III in boxes of sides of 1 Gpc/h, where $h$ is the reduced Hubble constant. This work uses these simulations to train and test our CNN.\footnote{The detailed document of the Quijote simulations can be accessed \href{https://quijote-simulations.readthedocs.io/en/latest/index.html}{here}. The simulations data can be accessed through \texttt{binder}, which can be accessed from \href{https://quijote-simulations.readthedocs.io/en/latest/access.html}{here}.} 

\begin{figure}
    \centering
        \includegraphics[width=0.85\linewidth]{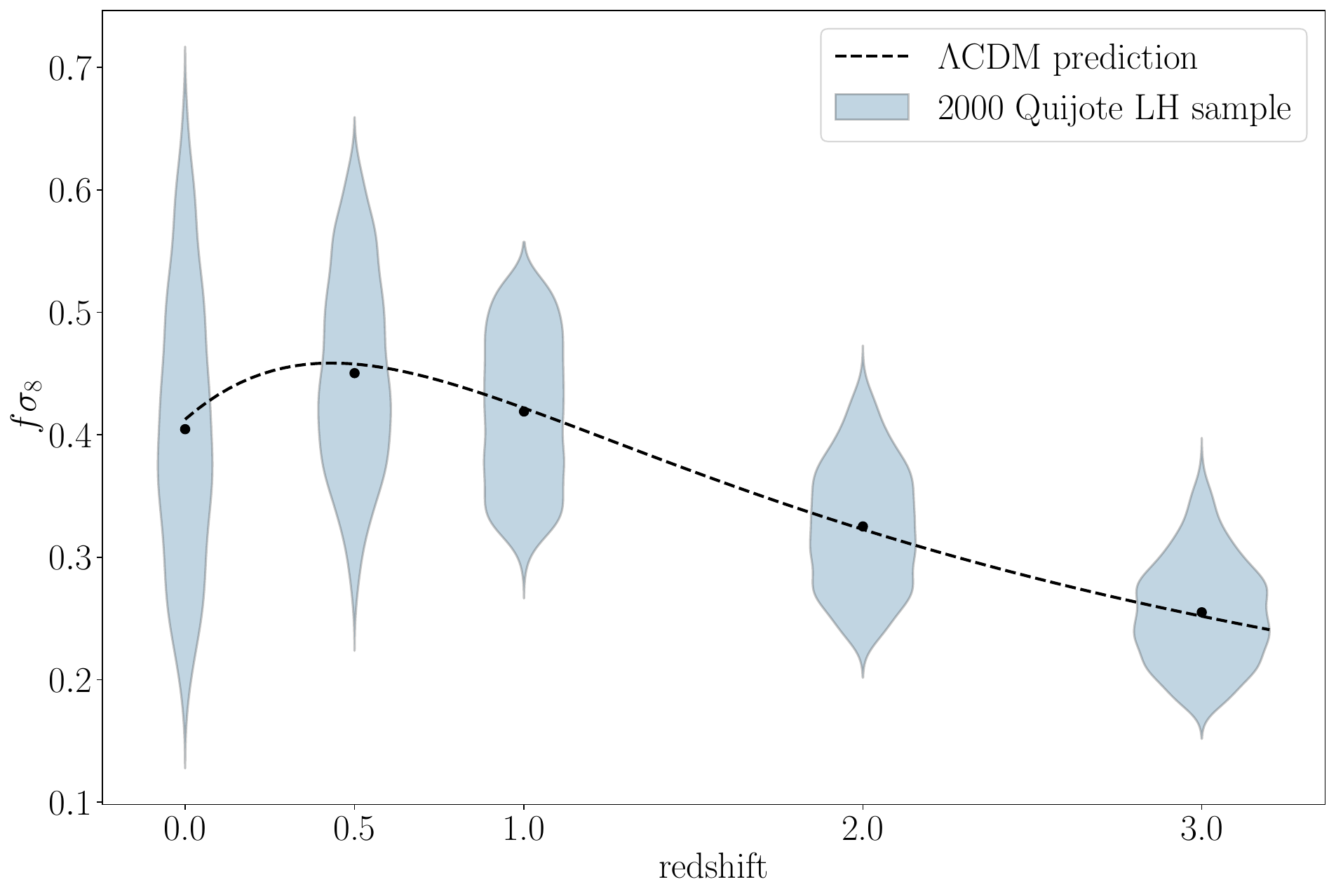}
    \caption{
    The distribution of the $f \sigma_8$ parameter values for Latin-hypercube realizations of the Quijote simulations at their corresponding redshifts. Black dots represent the mean value at each redshift, and the dashed curve shows the $\Lambda$CDM prediction.}
    \label{fig:fsigma8_dist}
\end{figure}

The Quijote simulations have $2000$ realizations with different combinations of cosmological parameters $(\Omega_m, \Omega_b, h, n_s, \sigma_8)$, chosen by the Latin-hypercube sampling within the ranges of 
\begin{eqnarray}
    \Omega_m &\in& [0.1, 0.5], \nonumber \\
    \Omega_b &\in& [0.03, 0.07], \nonumber \\
    h &\in& [0.5, 0.9], \nonumber \\
    n_s &\in& [0.8, 1.2], \nonumber \\
    \sigma_8 &\in& [0.6,1.0].
    \label{eq:cosmo_par}
\end{eqnarray}
All those simulations have different seeds of the initial condition. A single simulation contains $512^3$ dark matter particles and each realization has $5$ snapshots at redshifts $z=[3.0,\ 2.0,\ 1.0,\ 0.5,\ 0.0]$. In Fig.~\ref{fig:fsigma8_dist}, we show the $f \sigma_8$ distribution for 2000 realizations.

\subsection{The halo catalog}
\label{ssec:halo_catalog}

\begin{table}
  \centering
  \begin{tabular}{ccc}
    \hline \hline
    $z$ & relative number & $n_{\rm halo}$ \\ 
    & & $[(\mathrm{Gpc} h^{-1})^{-3}]$ \\
    \hline
    0.5 & 2 & $8.0 \times 10^3$ \\
    1.0 & 4 & $1.6 \times 10^4$ \\
    2.0 & 3 & $1.2 \times 10^4$ \\
    3.0 & 2 & $8.0 \times 10^3$ \\
    \hline \hline
  \end{tabular}
  \caption{
  Specifications for the mock simulation. It includes the relative number, which represents the expected number of halos based on an observation akin to the \textit{Euclid} survey, and  $n_{\rm halo}$ denotes the total number of halos incorporated within the simulation box, after the random selection process.
  }
  \label{tb:halo_number}
\end{table}

The Quijote simulations have a halo catalogue built by a Friends-Of-Friends (FOF) algorithm, where each halo has at least $20$ dark matter particles. Therefore, the lower limit of the halo mass is $\mathcal{O}(10^{12})$ - $\mathcal{O}(10^{13}) \ [M_\odot / h]$ (the value varies with the values of the cosmological parameters). 

We use the snapshots of $z=[0.5, 1.0, 2.0, 3.0]$ from the Quijote simulations.
Then, we assume the redshift halo distribution that roughly corresponds to the \textit{Euclid} survey and randomly pick up halos such that the redshift distribution matches the relative numbers in Table~\ref{tb:halo_number}. This table shows the mean of the comoving number density of the halos in each redshift bin over 2,000 realizations after after the selection process.

\subsection{Data for Machine Learning}
\label{ssec:make_image}

The architecture of our CNN will be discussed in Sec. \ref{ssec:MLarchitecture}. For the training part, we use the dark matter or halo distribution, and we build images for our CNN as follows: first, we define the $40^3$ grids in a simulation box, while this size of the grid corresponds to $k = 0.25\ \hmpc$ in Fourier space, and then we redistribute dark matter particles or halos to the cells by Nearest Gridding Point. As a result, we get the 3D images whose size is $40^3$, and each voxel value is the density of dark matter or halo in each cell. 
Note that this grid size is larger than the scale of the redshift distortion. In fact, we have tested our CNN using images in redshift space and found that the results do not change from those obtained using images in real space.

And then, for comparison, we train and test a neural network (NN) by 
the Legendre multipoles of power spectrum 
which is the two-point statistics in the redshift space defined as
\begin{align}
\label{eq:Pkl}
P_l (k) &= \frac{2l + 1}{2} \int^{1}_{-1} P(k, \mu) L_l (\mu) d \mu ,
\end{align}
where $P(k, \mu)$ is
the anisotropic power spectrum in redshift space,
and $L_l(\mu)$ is the Legendre polynomial of order $l$.
In this work, we use the expansion coefficient $P_0(k)$, $P_2(k)$, and $P_4(k)$ which are already calculated for the dark matter distribution in the simulations we introduce and are publicly available. As the input to our NN, we use the $P_0(k)$, $P_2(k)$, and $P_4(k)$ for $0.089 < k\ [\hmpc] < 0.25$, linearly separated in 39 bins.

The Quijote simulations have $2000$ realizations, so we use $1500$ simulations as training data, $100$ as validation data, and $400$ as test data for both, the CNN and the NN.

%
\section{Machine learning based cosmological parameter inference}
\label{sec:density}
%

 \begin{table}[h]
   \centering\setlength\tabcolsep{0pt}
    \begin{tabular*}{\linewidth}{@{\extracolsep{\fill}}r  c  c  c  c  c}
     \hline \hline
      method & \multicolumn{2}{c}{CNN} & PS-ML & Likelihood \\ \hline
      tracer & DM & halo & DM & DM \\ \hline
     $z = 0.5$ & 3.8  & 4.5 (3.9) & 2.3 & 7.9 \\
         $1.0$ & 2.3  & 2.5 (2.2) & 2.0 & 5.7 \\
         $2.0$ & 1.2  & 1.7 (1.1) & 2.7 & 3.9 \\
         $3.0$  & 0.74 & 1.2 (1.2) & 2.9 & 3.3 \\
     \hline \hline
   \end{tabular*}
   \rightline{($\times 10^{-2}$)}
   \caption{
   The predicted errors on the $f\sigma_8$ as derived by different methods and tracers.
   For the method, CNN, PS-ML, and Likelihood stand for convolutional neural networks, the power spectrum analysed by machine learning, and the maximum likelihood analysis, respectively.
   The values for ``halo" are obtained from the halo data, resampled from the parent halo to match the redshift distribution of our simulated observation (see Table \ref{tb:halo_number}) for CNN, while those in parenthesis are from the parent halo sample as a reference.
   }
   \label{tb:error_summary}
 \end{table}

In this section, we describe the ML approach to extract cosmological parameters from the images of the density field or the measured power spectra using the Quijote simulation. Furthermore, we employ maximum likelihood analysis to estimate the value of $f \sigma_8$, which we compare with the CNN approach. We first provide a basic structure of the ML architecture, specifically optimized for analysing images and power spectrum, and the method of the maximum likelihood analysis.
We then show the results of the comparison of different methodologies. 
We first compare the result of the CNN on the dark matter density field images to those on the halo density field images (Sec.~\ref{ssec:cnn_dm_halo}). This will directly compare the ability to constrain 
$f\sigma_8$ between dark matter and halo. 
We then compare the result of the CNN on the dark matter component with the ML-based dark matter power spectrum result and the maximum likelihood analysis (Sec.~\ref{ssec:dm_cnn_pk}). 

The predicted errors on the $f\sigma_8$ quantity as derived by different methods are summarized in Table~\ref{tb:error_summary}. In what follows, we discuss in detail how the analysis is performed for each method and how we perform the measurements in each case.

\subsection{ML architecture}
\label{ssec:MLarchitecture}

\begin{table}[h]
\centering
  \begin{tabular}{ c | c  c }
    \hline \hline
    \ & Layer & Output map size \\ \hline
    1 & Input & $40 \times 40 \times 40 \times 1$ \\
    2 & $3 \times 3 \times 3$ convolution & $38 \times 38 \times 38 \times 2$ \\

    3 & BatchNorm3d & $38 \times 38 \times 38 \times 2$ \\
    4 & $2 \times 2 \times 2$ MaxPool & $19 \times 19 \times 19 \times 2$ \\
    5 & $2 \times 2 \times 2$ convolution & $18 \times 18 \times 18 \times 64$ \\

    6 & BatchNorm3D & $18 \times 18 \times 18 \times 64$ \\
    7 & $2 \times 2 \times 2$ MaxPool & $ 9 \times 9 \times 9 \times 64$ \\
    8 & $3 \times 3 \times 3$ convolution & $7 \times 7 \times 7 \times 64$ \\

    9 & $3 \times 3 \times 3$ convolution & $5 \times 5 \times 5 \times 64$ \\

    10 & $2 \times 2 \times 2$ convolution & $4 \times 4 \times 4 \times 128$ \\

    11 & BatchNorm3d & $4 \times 4 \times 4 \times 128$ \\
    12 & Flatten & $8192 (= 4^3 \times 128)$ \\
    13 & FullyConnected & $512$ \\

    14 & FullyConnected & $256$ \\

    15 & FullyConnected & $1$ \\ \hline \hline
  \end{tabular}
  \caption{Our CNN architecture. In all convolutional layers, stride $=1$ and padding is not applied. Output map size corresponds to (height, width, depth, and channel). After each convolution layer and FullyConnected layer, except for the last layer, we apply the ReLU as the activation function. The total number of trainable parameters is 5,345,341.}
  \label{tb:CNN_architecture}
\end{table}

\begin{table}[h]
\centering
  \begin{tabular}{ c | c  c }
    \hline \hline
    \ & Layer & Output size \\ \hline
    1 & Input & $3 \times 39$ \\
    2 & FullyConnected & $512$ \\
    3 & FullyConnected & $512$ \\
    4 & FullyConnected & $512$ \\
    5 & FullyConnected & $512$ \\
    6 & FullyConnected & $1$ \\ \hline \hline
    \end{tabular}
  \caption{Our NN architecture. After each FullyConnected layer, except for the last layer, we apply the dropout layer with rates of $0.1$. The total number of trainable parameters is $848,897$.}
  \label{tb:NN_architecture}
\end{table}

This work uses the 3-dimensional CNN to analyse the images and an NN for the power spectrum. We used the publicly available platform PyTorch~\cite{pytorch} to construct our CNN and NN and the architecture based on Ref.~\cite{Lazanu:2021tdl}; however, note that some hyperparameters are different from those of that work, as the size of input data is different. In Tables~\ref{tb:CNN_architecture} and \ref{tb:NN_architecture}, we show the architecture of our CNN and NN, while it should be noted that for the activation function, we apply the ReLU after each convolution layer and FullyConnected layer except for the last layer. Also, our CNN predicts the value of $f \sigma_8 (z)$ from the $40 \times 40 \times40$ image of dark matter or halo distribution introduced in Sec.~\ref{ssec:make_image}.

Here, we also use mini-batch learning, and when we choose $N_b$ as the batch size, we randomly divide the training data into groups, and each group has $N_b$ training data. This group is called a mini-batch, and then the average value of the loss function in the mini-batch is used to update the trainable parameters. With trial and error, we determined that the optimal batch size is $16$; however, if we make the batch size unity, the loss value does not converge because we use batch normalization in our CNN.

As our main loss function, we use the Mean Squared Error (MSE)
\begin{equation}
    \mathcal{L} = \frac{1}{N_b} \sum_{i=1}^{N_b} (y_i - \hat{y}_i)^2 ,
\end{equation}
where $N_b$ is the batch size, $y_i$ is the predicted value of $f \sigma_8$ from our CNN for the $i$-th data in the mini-batch data, and $\hat{y}_i$ is the ground-true value, i.e. the correct $f \sigma_8$ value.

In updating the trainable parameters, we use the Adam optimizer~\cite{AdamOptimizer}, which is defined as \texttt{torch.optim.Adam()} in PyTorch. We use $5 \times 10^{-7}$ and 0.1 as the value of lr and weight\_decay, which are the arguments of \texttt{torch.optim.Adam()}, respectively.
The batch normalization~\cite{BatchNorm} is used in our CNN as we have found that this improves the learning efficiency. This is defined as \texttt{torch.nn.BatchNorm3d()} in PyTorch, and we use the default values for all parameters.

Our NN predicts the value of $f \sigma_8$ from the Legendre expanded power spectrum.
We use the $P_0(k)$, $P_2(k)$, and $P_4(k)$ for $k \le 0.25$ $\hmpc$, so the size of input data to our NN is $3\ (P_0,\ P_2,\ \mathrm{and}\ P_4) \times 39\ (\mathrm{the\ number\ of
\ k\textrm{-}bins})$. Furthermore, we apply the dropout layer with the rate of 0.1 after each Fully Connected layer, except for the 6th layer, see Table \ref{tb:NN_architecture}. As the loss function and the optimizer of the trainable parameters, we use the MSE and the Adam optimizer where lr$=5 \times 10^{-6}$ and weight\_decay$=0$. In addition, we apply $N_b=16$ as the batch size in training.

%
\subsection{Parameter Inference by the Maximum Likelihood}
\label{ssec:likelihood_inference}
%

For the comparison between machine learning and conventional approaches, we estimate $f \sigma_8$ using maximum likelihood for the dark matter density field. Here, we consider the likelihood below:
\begin{align}
    -2 \ln L = \sum_{i,j}\  &[ P^{\rm model}_i - P^{\rm data}_i] \nonumber \\
    &\times \mathrm{Cov}^{-1}_{i,j} \nonumber \\
    &\times [ P^{\rm model}_j - P^{\rm data}_j].
\end{align}
Here, $P^{X}_i$, where $X$ is either `model' or `data', represents the concatenated Legendre multipoles $[P_0(k), P_2(k), P_4 (k)]$ of the power spectrum as defined in Eq.(\ref{eq:Pkl}). We restrict our analysis to $k < 0.2\ h \mathrm{Mpc}^{-1}$, which is justified by the maximum wave number applicable for the EFT power spectrum utilized \citep{2022JCAP...11..038N}. Each multipole comprises 31 bins of the wave number,  resulting in $i$ ranging from 1 to 93 ($=3 \times 31$).

Also, $P^{\rm data}$ denotes the power spectrum derived from the Quijote Latin Hypercube (LH) simulation dataset, where cosmological parameters are sampled using Latin-Hypercube sampling. $\mathrm{Cov}$ represents the covariance matrix of $P^{\rm data}$. To estimate the covariance matrix, we use 15,000 realizations from the `Fid' simulation dataset of Quijote, where the cosmological parameters are [$\Omega_m$, $\Omega_b$, $h$, $n_s$, $\sigma_8$] = [$0.3175, 0.049, 0.6711, 0.9624, 0.834$]. Each realization is generated with a different random seed for its initial condition, thus accounting for cosmic variance in the covariance estimation.

Finally, $P^{\rm model}_i$ corresponds to the power spectrum calculated following Eqs.(\ref{eq:EFT_pk}) and (\ref{eq:Pkl}). To compute it, we use \texttt{FOLPS-nu}\footnote{https://github.com/henoriega/FOLPS-nu}, which computes the Legendre multipoles of the EFT power spectrum \citep{2022JCAP...11..038N}, and \texttt{CLASS} \citep{Lesgourgues:2011re} for the linear power spectrum. Here, we focus on an unbiased tracer (dark matter). Specifically, we consider the five cosmological parameters [$\Omega_m$, $\Omega_b$, $h$, $n_s$, $\sigma_8$] and four nuisance parameters [$\alpha_0$, $\alpha_2$, $\alpha^{shot}_0$, $\alpha^{shot}_2$] in Eq.\eqref{eq:EFT_pk}, setting both $\alpha_4$ and $\alpha_6$ to zero following \cite{2022JCAP...11..038N}. We randomly select 400 realizations from the LH dataset, compute $P^{\rm data}$ for each realization, and fit the aforementioned $5+4$ parameters to minimize $- \ln L$. For the minimization of $- \ln L$ (equivalent to maximizing the likelihood), we use \texttt{scipy.optimize.minimize()}, implemented in \texttt{scipy} \citep{2020SciPy-NMeth}. Subsequently, we calculate $f\sigma_8$ from the fitted parameters in each realization following Eqs.~(\ref{eq:delta_m})-(\ref{eq:calc_fs8}).

%
\subsection{CNN results on dark matter and halo}
\label{ssec:cnn_dm_halo}
%

\begin{figure*}
\centering
\begin{tabular}{cc}
\hspace{-0.4in} \includegraphics[width=4.5in]{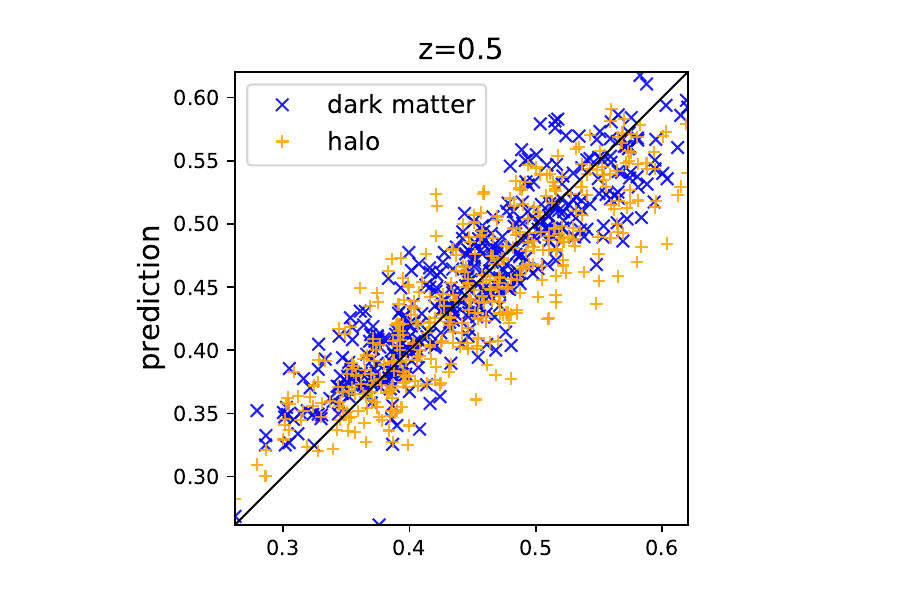} & 
\hspace{-1.4in} \includegraphics[width=4.5in]{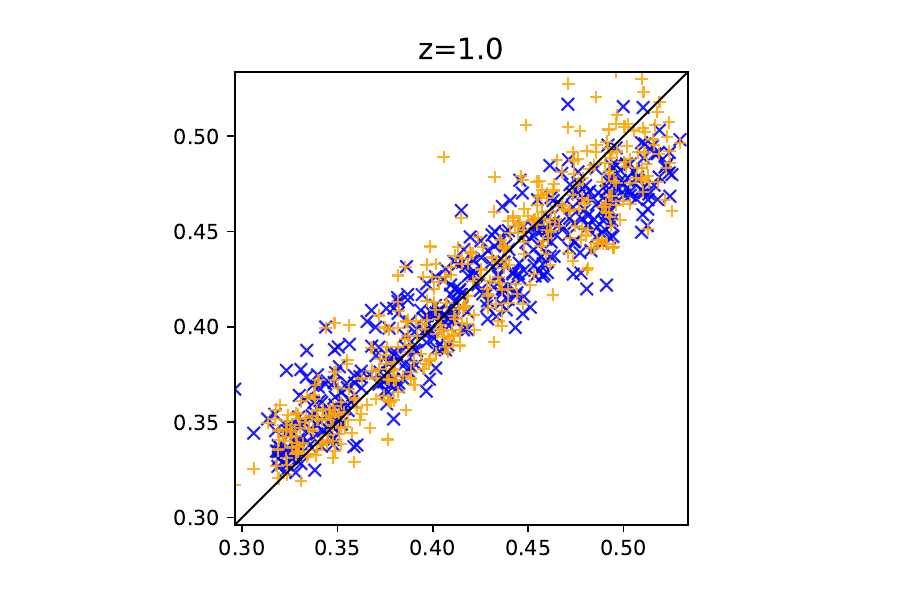} \\
\hspace{-0.4in} \includegraphics[width=4.5in]{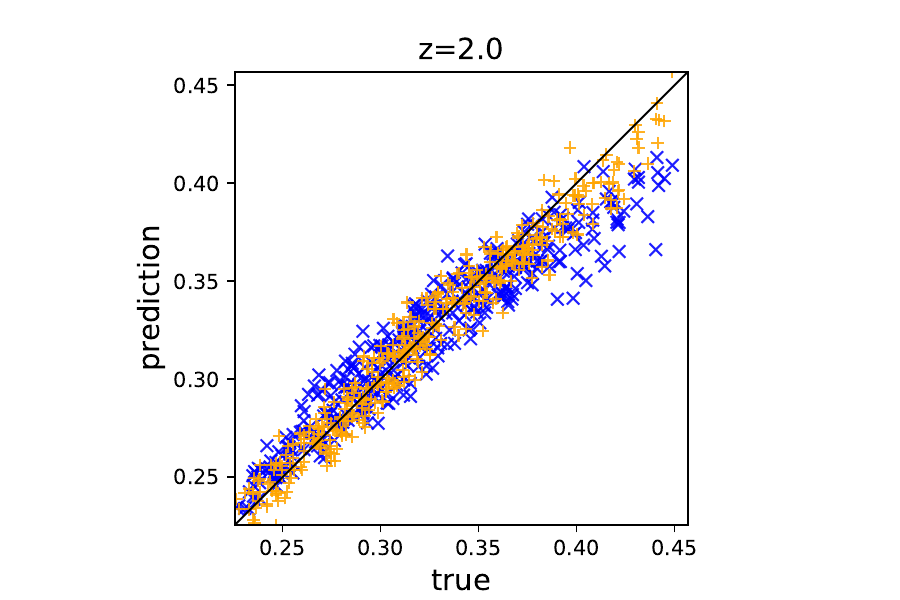} & 
\hspace{-1.4in} \includegraphics[width=4.5in]{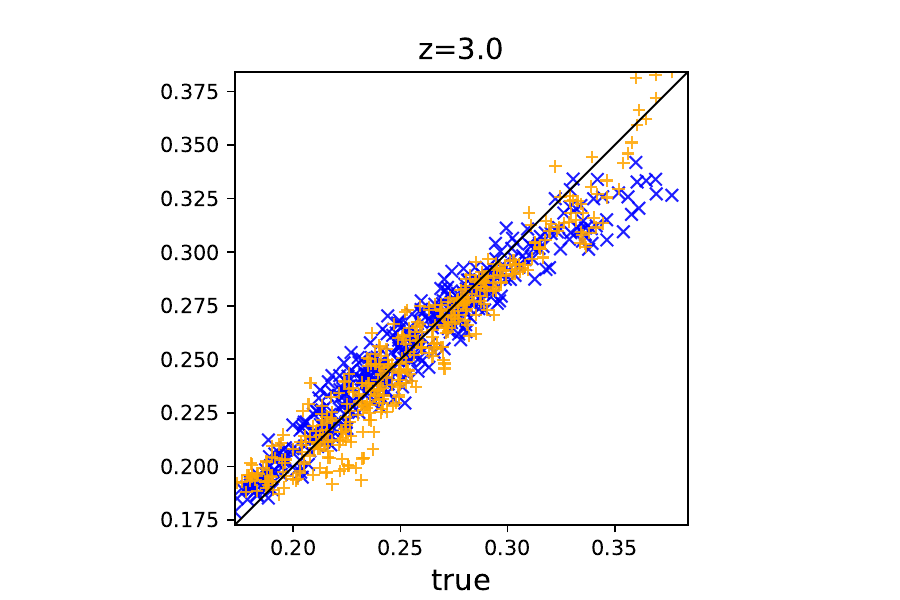}
\end{tabular}
\caption{The results of our CNN for dark matter images (blue) and halo images, including all halo (orange) for each redshift. The horizontal axis shows the true value of $f \sigma_8$, and the vertical axis shows our CNN's predicted value of $f \sigma_8$.
}
\label{fig:CNN_scatter}
\end{figure*}

\begin{figure*}[ht]
\begin{center}
\begin{tabular}{cc}
\hspace{-0.2in} \includegraphics[width=3.5in]{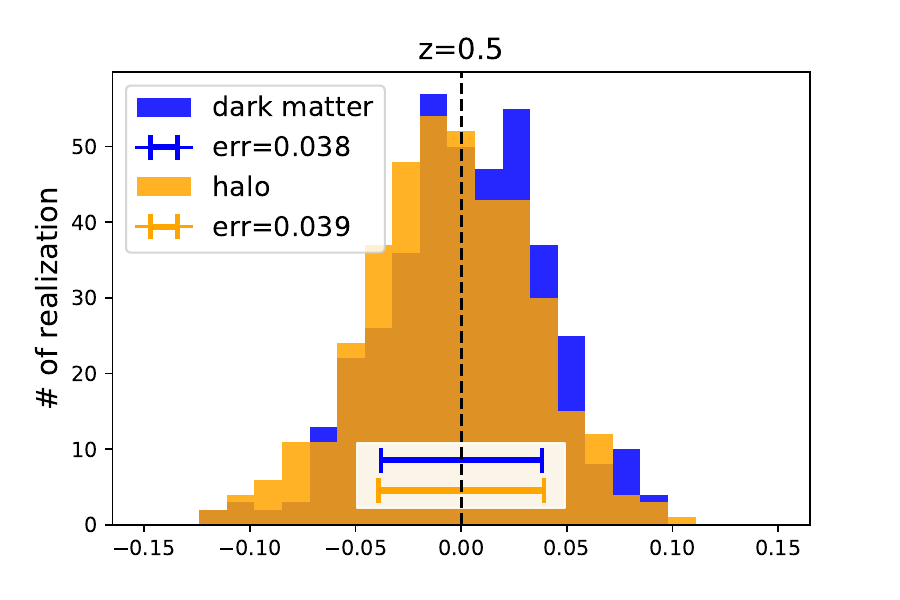} & \hspace{-0.5in} \includegraphics[width=3.5in]{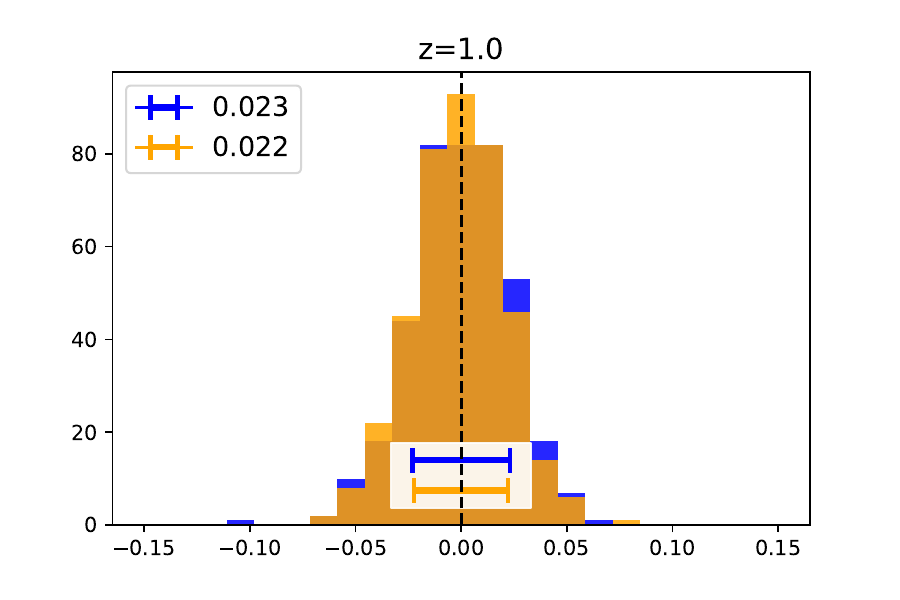} \\
\hspace{-0.2in} \includegraphics[width=3.5in]{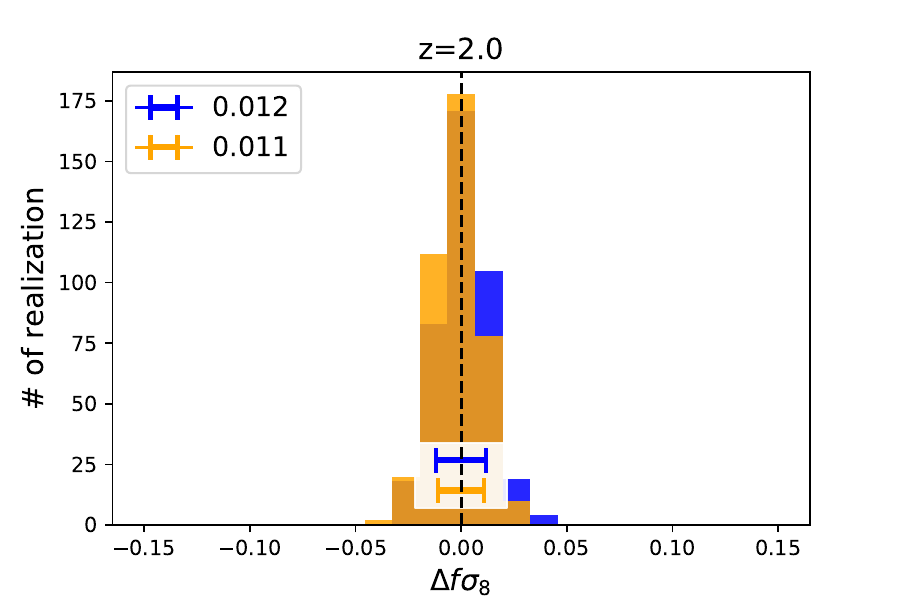} & \hspace{-0.5in} \includegraphics[width=3.5in]{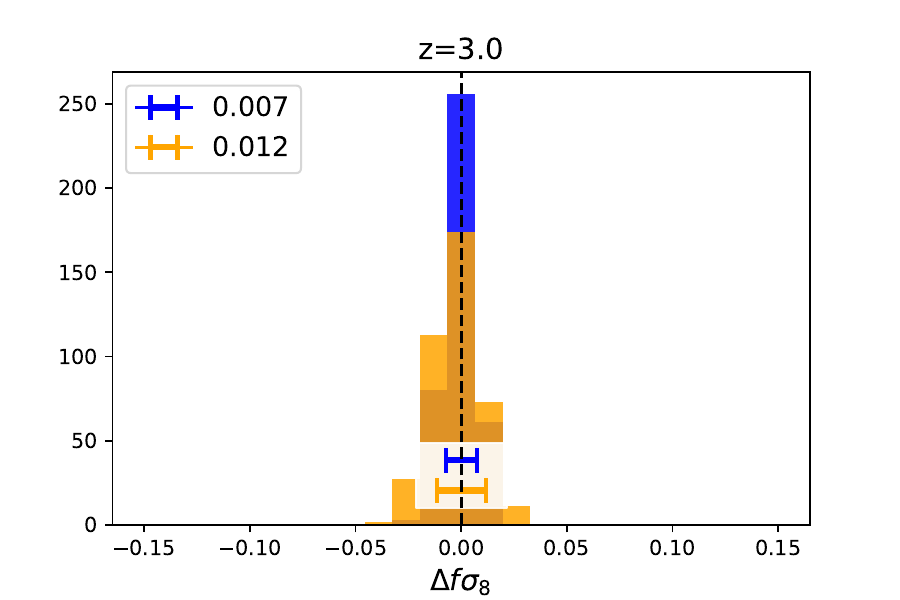}
\end{tabular}
\end{center}
\caption{The histograms of the CNN prediction for dark matter images (blue) and halo images (orange) for each redshift bin. The horizontal axis corresponds to the value of $\Delta f \sigma_8$ predicted by our CNN, and the vertical axis shows the number of test images for each $\Delta f \sigma_8$. The error bars show the standard deviation of $\Delta f \sigma_8$; its value is shown in the legend.
It can be noticed that all histograms are centred around zero, indicating that our approach successfully constrains $f \sigma_8$. 
}
\label{fig:CNN_hist_dm_halo}
\end{figure*}

\begin{figure*}[ht]
\begin{center}
\begin{tabular}{cc}
\hspace{-0.2in} \includegraphics[width=3.5in]{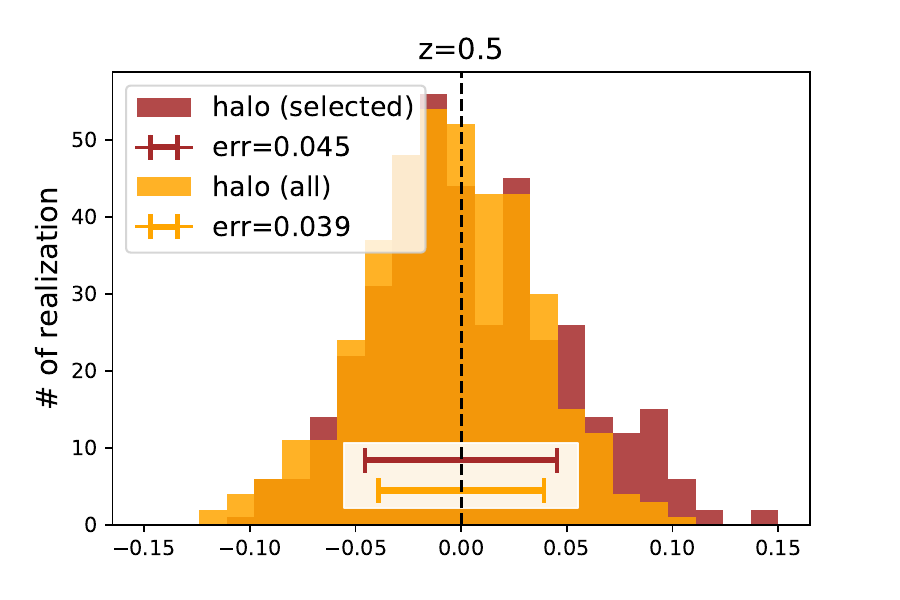} & \hspace{-0.5in} \includegraphics[width=3.5in]{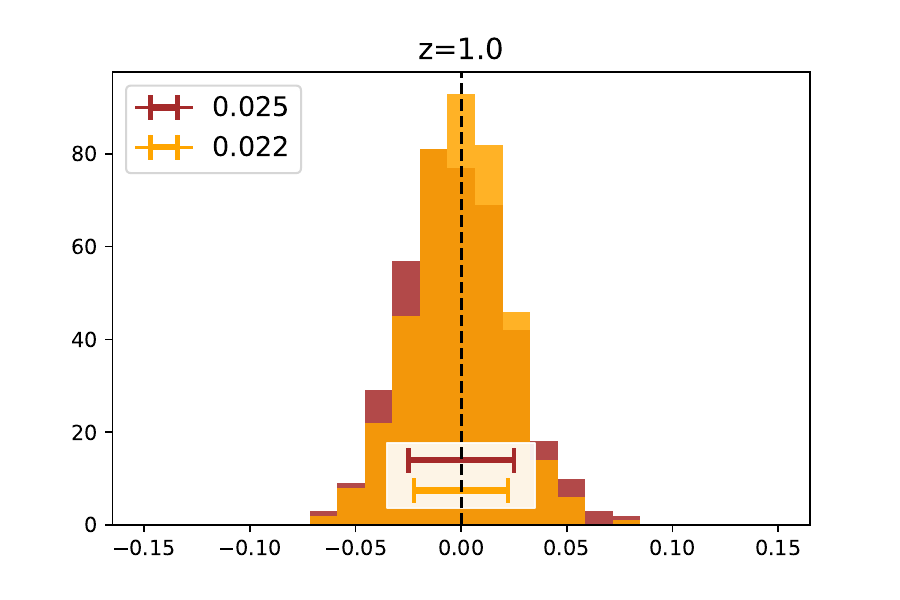} \\
\hspace{-0.2in} \includegraphics[width=3.5in]{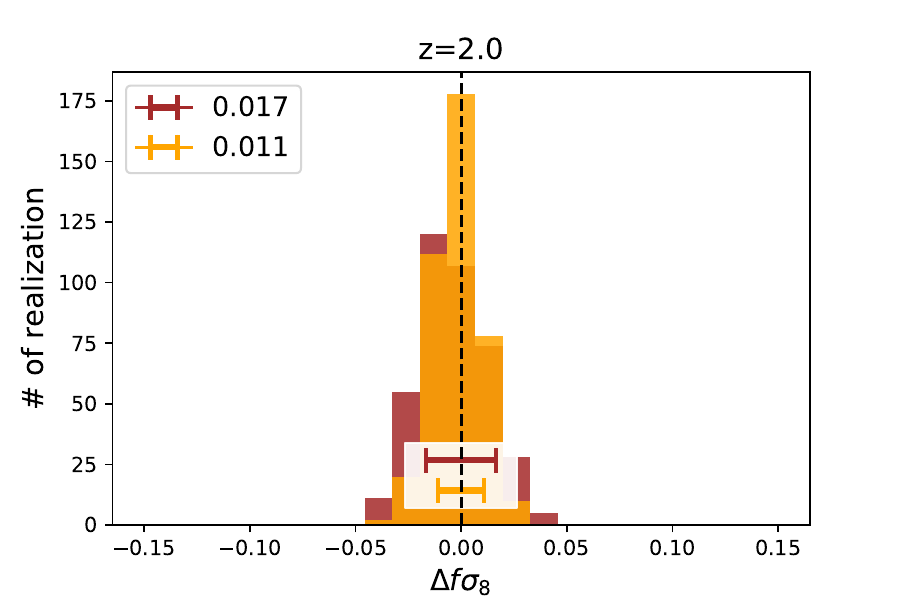} & \hspace{-0.5in} \includegraphics[width=3.5in]{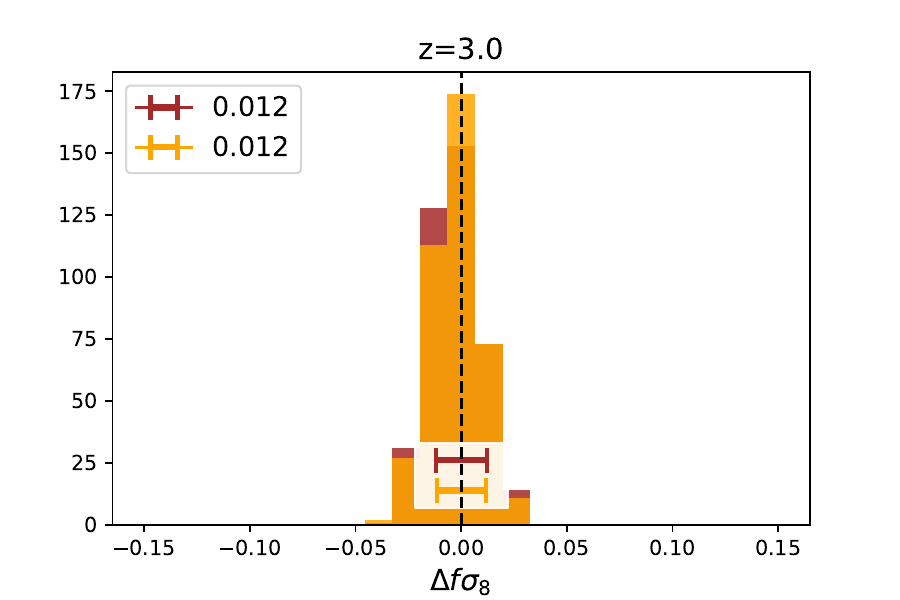}
\end{tabular}
\end{center}
\caption{The results of our CNN for images including all halos (orange) and randomly selected halos (brown) for each redshift bin. The horizontal axis corresponds to the value of $\Delta f \sigma_8$ predicted by our CNN, and the vertical axis shows the number of the test images for each $\Delta f \sigma_8$. The error bars show the standard deviation of $\Delta f \sigma_8$, and its value is shown in the legend.
}
\label{fig:CNN_hist_all_selected_halo}
\end{figure*}

In this subsection, we compare the results from our CNN for the images of dark matter distribution and the halo distribution.

During the training phase, our CNN undergoes $100$ epochs, each representing a complete iteration of the CNN learning process using the entire training dataset. Then, in the test phase, we use the CNN model, which minimizes the value of the loss function for the validation data. 

In this work, the errors are estimated by the standard deviation of the vector of $\Delta f \sigma_{8, i} = (y_{\mathrm{pred}, i} - y_{\mathrm{true}, i})$, where $y_{\mathrm{pred}, i}$ and $y_{\mathrm{true}, i}$ are the CNN prediction and the ground truth for the $i$-th test data, respectively. 

To probe the validity of this estimation, we perform the following test:
 \begin{itemize}
     \item We make a $30 \times 30$ grid of parameters for $\Omega_{m} \in [0.1, 0.5]$ and $\sigma_8 \in [0.6, 1.0]$.
     \item For each redshift, we evaluate the $f \sigma_8$ in the whole grid, and then, we pick 400 grid points (the same number of the test data for our CNN) from them.
     \item We draw predictions for $f \sigma_8$ based on a normal distribution with mean given by the value at each point and with error $\sigma$. This is justified by considering a large number of simulations and the central limit theorem. 
     \item We compare the predicted error from the standard deviation of 
     $\Delta f \sigma_{8, i}$ vector with the fiducial error assumed in the previous step.
 \end{itemize}

Figure~\ref{fig:CNN_scatter} 
shows the relation between true and predicted $f\sigma_8$
for our CNN results, respectively for dark matter (blue) and halo (orange) test images. Figure~\ref{fig:CNN_hist_dm_halo} displays histograms of $\Delta f \sigma_8$ for the test data, allowing us to compare errors between the dark matter and halo images. As shown in Figure~\ref{fig:CNN_hist_dm_halo}, all histograms are centred around zero, indicating the effectiveness of our approach in constraining $f \sigma_8.$ 

Here, for comparison purposes, we use all the halos when building the images (without disregarding any halo). By comparing the results for dark matter and halo images, we find that the errors are mostly comparable across all redshift bins, while the results for the dark matter images are better at $z=3.0$. The error decreases as the redshift increases for both dark matter and halo images. One possibility is that non-linearities make it more challenging to extract information from the matter distribution at low redshifts. At $z=3.0$,  the error for the halo images is larger than that for the dark matter and comparable to the error for $z=2.0$ halo images. We can attribute this to shot noise because the number of halos in the simulation at $z=3.0$ is less than one-tenth of the number at other redshifts when considering all halos.

As a side note, when we test our CNN with the dark matter images of $64^3$ grids, corresponding to $k=0.4\ \hmpc$, we observe slightly improved errors by approximately $15\%$ across all the redshifts we consider. Given the configuration of the Quijote simulation we utilize, the Poisson shot noise dominates the dark matter power spectrum at $k = \mathcal{O}(1)\ \hmpc$. It is probable that the CNN performs well in analyzing these scales.

Next, we investigate the effect of the random selection of the halos. Instead of the images including all halos, 
we reduce the number of halos so that its redshift distribution follows a realistic observation (the ratio is
shown in Table~\ref{tb:halo_number}) and use them to train and test our CNN. The results are shown in Fig.~\ref{fig:CNN_hist_all_selected_halo}. Even when we use only randomly selected halos, we can see the redshift dependence of the error as in the all-halo case. The errors of the selected halo case are more significant than the ones for all the halos, but this is reasonable because the images of the selected halos lose the information compared to the images including all halos.

It is important to mention that, by assuming a redshift-halo distribution that matches the one of surveys like \textit{Euclid}, our intention is to make an analysis with Stage IV-like (realistic) data. However, to make a direct comparison with a real catalog, additional criteria must be addressed. An initial strategy could involve training a CNN to analyze observable features within a galaxy survey, such as a luminosity-limited galaxy sample, which inherently contains noise. Subsequently, the CNN could extract $f\sigma_8$ from these observations.

In a simulation suite like Quijote, where the matter density varies, the minimum mass of halos fluctuates according to the cosmological parameters. Another approach could be to train on a halo catalog with a minimum luminosity matching the one expected from \textit{Euclid}. These could be interesting ways of complementing the realistic-like catalog analysis, but this is beyond the scope of this work.

We additionally performed some tests in the CNN architecture by exploring different loss functions to corroborate our results. We studied the CNN behaviour with MSE Loss, MAE (Mean Absolute Error) Loss, Hubber Loss, and the moments network Loss 
(LFI - Likelihood Free Inference)~\cite{villaescusa2022camels, jeffrey2020solving}. We observe a slight improvement in error with the LFI loss at low redshift. However, when evaluating the loss curve, we find that the MSE loss is more stable, so we chose to implement it for our final results. See Appendix A for more details.

%
\subsection{ML-based power spectrum analysis on dark matter}
\label{ssec:dm_cnn_pk}
%

\begin{figure*}[ht]
\begin{center}
\begin{tabular}{cc}
\hspace{-0.2in} \includegraphics[width=3.5in]{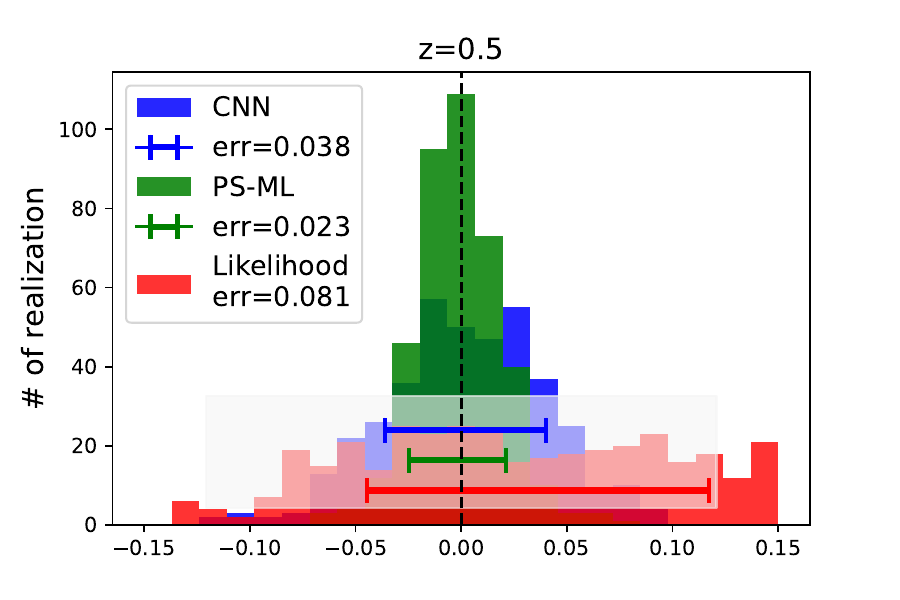} & \hspace{-0.5in} \includegraphics[width=3.5in]{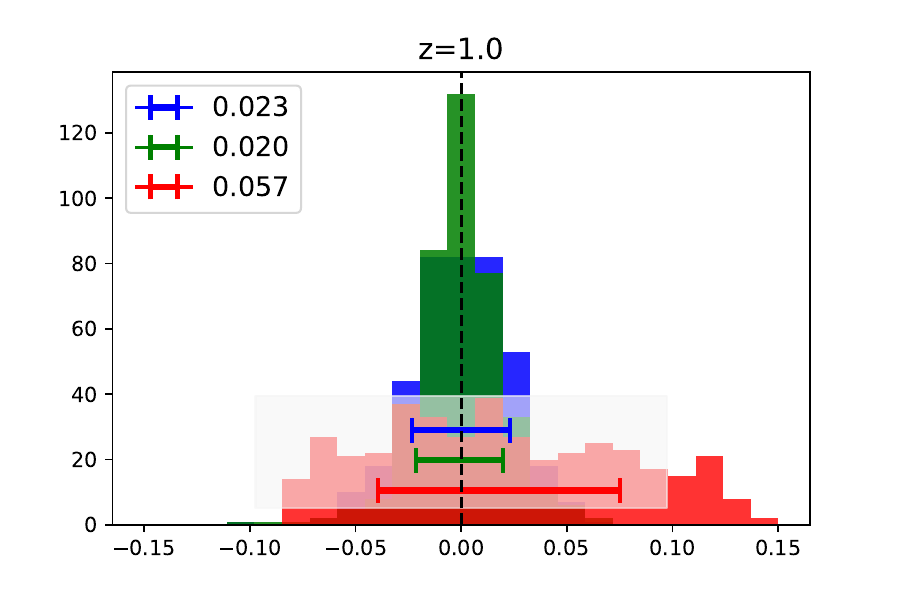} \\
\hspace{-0.2in} \includegraphics[width=3.5in]{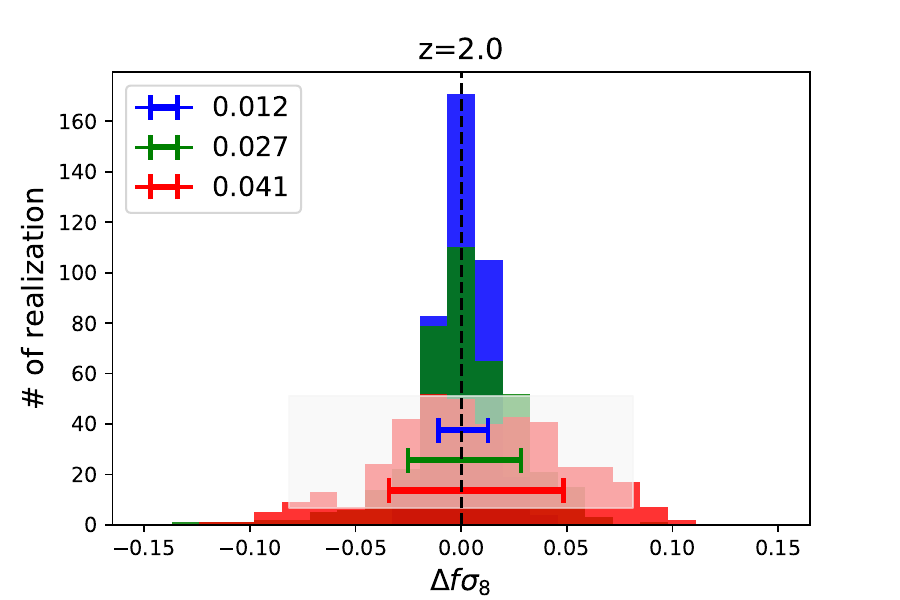} & \hspace{-0.5in} \includegraphics[width=3.5in]{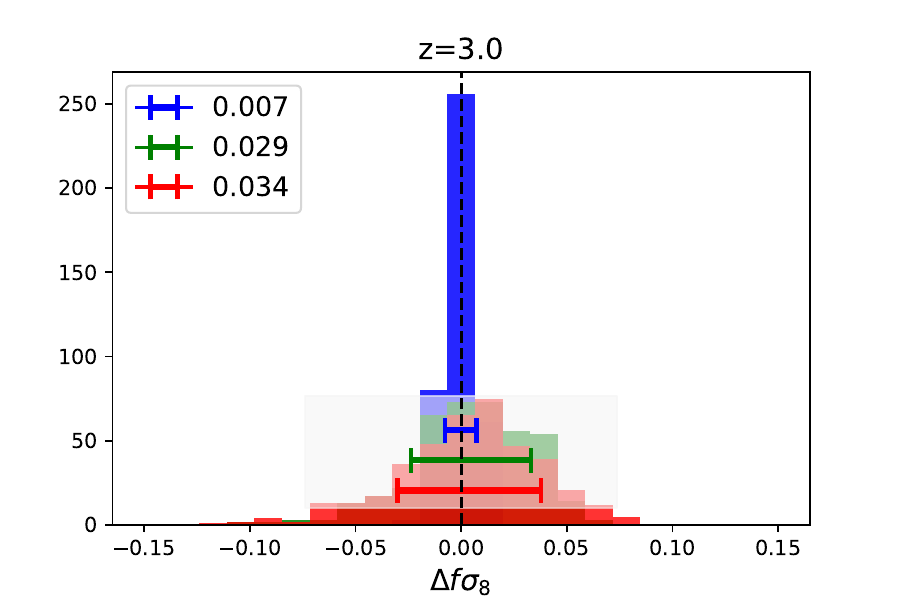}
\end{tabular}
\end{center}
\caption{
The results of our CNN for dark matter images (blue), PS-ML (green), and likelihood (red) for each redshift. The horizontal axis corresponds to the value of $\Delta f \sigma_8$ predicted by our CNN, PS-ML, and likelihood, and the vertical axis shows the number of test images for each $\Delta f \sigma_8$. The error bars show the standard deviation of $\Delta f \sigma_8$ for each method. The values of these errors are shown in the legend.
}
\label{fig:CNN_NNpk}
\end{figure*}

Now, we compare the CNN results with the ones from the machine-learning-based power spectrum analysis (PS-ML) for dark matter. In this subsection, we use the NN whose architecture is defined in Table \ref{tb:NN_architecture} for the power spectrum analysis. Training and error evaluation processes are exactly the same for our CNN in Sec.~\ref{ssec:cnn_dm_halo}.

In Fig.~\ref{fig:CNN_NNpk}, we compare the results obtained by dark matter images (blue, same as the one in Fig.~\ref{fig:CNN_hist_dm_halo}) and PS-ML (green). 
We see that the constraints on $f\sigma_8$ from PS-ML are weaker at higher redshifts ($z\ge2.0$) while it goes opposite at low redshift $z\le1.0$. The results look puzzling, but the constraining power may depend on the architecture of the CNN/NN, as we discussed in section \ref{ssec:cnn_dm_halo}. Therefore, it is not straightforward to interpret the results, but one possibility is that in the highly non-linear regime, image recognition tends to fail to extract the characteristics of the data.

In Fig.~\ref{fig:CNN_NNpk}, we also show the results of the maximum likelihood analysis for dark matter images (red). The standard deviation of $\Delta f \sigma_8$ is represented by the red error bar. We estimate the $f \sigma_8$ values for 400 realizations from the Quijote LH dataset following Section~\ref{ssec:likelihood_inference}. For each realization, we obtain the predicted $f \sigma_8$ value and evaluate $\Delta f \sigma_8$ in a manner similar to the ML analysis.  It is evident that CNN and PS-ML outperform the likelihood analysis, indicating their ability to extract more information from the dark matter density field than the modeled power spectrum in Eq.~(\ref{eq:EFT_pk}). 

Additionally, we can see two features in the results obtained from the maximum likelihood. First, the maximum likelihood analysis shows redshift dependence similar to CNN. This dependence is likely caused by non-linearities at lower redshifts. At $z=3$, the maximum likelihood and the PS-ML show comparable results because the non-linearity is weaker at higher redshifts. The PS-ML directly fits the observed power spectra for the simulation data, including the non-linear effects, to derive the $f \sigma_8$ value, and the non-linearity has less effect on the PS-ML compared to the maximum likelihood method. Second, the maximum likelihood analysis shows a bias of $\Delta f \sigma_8$ at lower redshifts. This bias may be caused by the model-dependent property of the analysis. On the other hand, CNN and PS-ML do not show such bias, and it is one of the advantages of the ML-based analysis.

%
\subsection{Effect of the random seed}
\label{ssec:random_seed}

Finally, we discuss the effect of the random initial conditions. So far, we have used the Latin-hypercube Quijote simulations, which have different random seeds for the initial conditions. Therefore, the difference in these simulations is caused by the difference in its cosmological parameters and initial conditions. To discuss the effect of the second one, we use another dataset of Quijote simulations. The Quijote simulations have 15000 realizations as `Fid' simulations for the same cosmological parameter set and different random seeds for their initial conditions.

First, we pick up 7000 realizations from the `Fid' simulation dataset, accessible on the San Diego Cluster hosting the Quijote simulations data. Then, we build the 3D images of randomly selected halo distributions from these data following Sec.~\ref{ssec:make_image}. 
Then, we use our CNN model, which is already trained by the randomly selected halo images in Sec.~\ref{ssec:cnn_dm_halo}, and test it by the images from the `Fid' simulations. By doing this, we can evaluate the error only from random initial conditions. As a result, we find that the error for `Fid' realizations is $\mathcal{O} (10^{-3})$.

This is about one-fifth of the error in the results in the previous subsections, which should include both effects from the variations of cosmological parameters and the initial conditions. Therefore, we can conclude that the error in our CNN analysis is dominated by the variance of cosmological parameters arising from different cosmologies in each realization. 

\section{Discussions and conclusions\label{sec:conclusions}}

The growth rate of matter density perturbations, described by the bias-independent $f\sigma_8(z)$ quantity, is an important quantity in studying the large-scale structure of the Universe, as it allows us to probe the dynamic features of gravity. However, to extract $f\sigma_8(z)$ from observations made by galaxy surveys, two key assumptions have to be made: i) a fiducial cosmological model and ii) the modeling of the non-linear part of the power spectrum. From these two assumptions, the latter is particularly insidious. Currently, only purely phenomenological models exist in the literature, thus potentially biasing the measurements of $f\sigma_8(z)$ in ways that are not easy to correct a posteriori.

The primary objective of this work, is to implement a ML approach based on training CNNs on N-body simulations to extract $f\sigma_8(z)$, without the reliance on any specific modeling of the non-linear power spectrum. This could serve as an initial step towards the future development of a method for parameter inference in Stage IV surveys, but for this, some considerations, like systematic and baryonic effects, need to be taken into account. In a sense, we perform a likelihood-free extraction of the growth rate, as the CNN requires neither a likelihood nor any modeling of the power spectrum, by leveraging the ability of the neural network architecture to predict the growth $f\sigma_8(z)$ from images of the dark matter or halo distributions build from the N-body simulations, as discussed in Sec.~\ref{ssec:MLarchitecture}. In addition, we compare the results of the ML approaches with those obtained by NN trained on the multipoles of the power spectrum and by the maximum likelihood analysis with the modeled power spectrum.

Specifically, our work followed a multi-pronged approach to find the optimal way to extract the growth via an ML architecture. 
The principal results and comparisons between the various approaches are shown in Figs.~\ref{fig:CNN_scatter}, \ref{fig:CNN_hist_dm_halo}, \ref{fig:CNN_hist_all_selected_halo}, and \ref{fig:CNN_NNpk} 
, where we showed scatter plots and histograms comparing the CNN dark matter and halo reconstructions, and the CNN dark matter and PS-ML reconstructions
Table~\ref{tb:error_summary} shows the predicted errors from all approaches.

Overall, as can be seen in the figures above and Table~\ref{tb:error_summary}, we find that the ML architecture can predict quite accurately the value of the growth rate $f\sigma_8(z)$ (as seen in Fig.~\ref{fig:CNN_hist_dm_halo} where all histograms are centred around zero). At the same time, the reconstructed errors from the CNN are comparable within a factor of order unity between the different ML approaches based on the DM or halo catalogues. Additionally, the ML approaches show smaller errors compared to likelihood analysis using modeled power spectra.
However, we observe the redshift dependence of the CNN results, while the errors calculated by the ML-based power spectrum analysis do not largely depend on the redshift. To probe for the cause of this redshift behaviour of the error from the CNN, we tested several possibilities, which, while they might not individually constitute the main reason, each may contribute to some extent.
 
First, one plausible reason for the redshift trend of the CNN results 
might be that at low redshifts, the enhanced non-linearities might introduce more scatter in the observed values of $f\sigma_8$ by the CNN, something which might not be captured correctly by the PS-ML.

Second, another possible reason for the decreasing error of the CNN estimate might be because while the Quijote simulations are sampled well on the $(\Omega_m, \Omega_b, h, n_s, \sigma_8)$ from the Latin-Hypercube, this is not the case for $f\sigma_8(z)$, as already seen in Fig.~\ref{fig:fsigma8_dist}. 
This would, in effect, introduce a prior on $f\sigma_8(z)$ that decreases/tightens with redshift. See the blue “blobs” in Fig.~\ref{fig:fsigma8_dist} that indicate the scatter of $f\sigma_8(z)$, thus artificially training the CNN to expect a reducing value on the error with redshift (since the possible scatter of the parameters is limited by the prior). 
However, we should note that the width of the prior is generally greater than the error of the CNN. We conducted a test by artificially narrowing the range of the prior (making the $f\sigma_8(z)$ width, equal for all the considered redshift bins). After training and testing our architecture, no significant impact on the error estimates was found.

To investigate the correct sampling of $f\sigma_8(z)$ in the Quijote simulations compared to $(\Omega_m, \Omega_b, h, n_s, \sigma_8)$, two approaches could be considered. First, increasing the number of simulations used in training could broaden the prior and address the observed redshift-dependence on the error in Fig.~\ref{fig:fsigma8_dist}. Alternatively, directly sampling within the $f\sigma_8(z)$ space, possibly facilitated by an N-body emulator, could be explored. This would require additional N-body simulations to potentially enhance resolution. However, due to the limited availability of Quijote simulations, leveraging an emulator to artificially augment the dataset is proposed as a means to address this limitation and potentially mitigate any observed trends in $f\sigma_8$. Nonetheless, conducting these additional tests is deemed beyond the scope of this study as we primarily aim to highlight our methodology.

Third, the training/testing approach of the CNN also has some stochasticity/randomness as different runs produce different errors. Thus, we tested this by running the testing phase with different random seeds and calculated the mean and standard deviation of the CNN error. In this case, we have found that the mean error is consistent with the values in Table~\ref{tb:error_summary} and the standard deviation of the error (i.e. the ``error of the error”) is much smaller. Thus, this reason could only contribute to a small extent to the redshift behaviour of the CNN error.

Finally, we also performed several more tests; for example, we examined the effect of the random selection of the halos by making images that have a part of halos to reproduce the relative number shown in Table~\ref{tb:halo_number} and then used them to train and test our CNN. Doing that, we found the same redshift dependence of the error as in the previous case. Second, we also explored different loss functions, including the MSE, MAE, Hubber Loss, and LFI loss functions. Overall, we found a similar decrease in the error with redshift. Third, we investigated the effect of the random initial conditions by using another dataset of Quijote simulations. Doing so, we find that the error contribution is only one-fifth of the total error. Thus, we conclude that the difference in the cosmological parameters dominates the error in our analysis.

We successfully trained and validated CNNs to predict $f\sigma_8(z)$, using various input data sources, such as dark matter density fields and halo catalogs. But the applicability of the trained CNNs is somehow limited when talking about its implementation to real data. Therefore, this could be taken only as a first step for developing ML-based methods for cosmological parameter inference from observations. This approach may require additional factors into consideration, including the systematic and baryonic effects, as well as the fact that in real galaxy surveys, the positions of halos are known within a certain degree of uncertainty, so the measurements of $f\sigma_8$ in these surveys carry uncertainties proportional to their inherent noise characteristics.
There were other attempts to perform parameter inference from simulations, for instance \cite{pandey2023sensitivity}, where $\Omega_m$ and $\sigma_8$ are inferred directly from the power spectrum multipoles, whereas in this study we employ the raw image of the density field. Certainly, if the Legendre polynomial power spectrum decomposition were available without the bias term, one could compute $\sigma_8$ directly. Furthermore, in the presence of a dark matter density field, it is feasible to integrate the variance of the matter density within 8 Mpc shells to derive $\sigma_8$. For example see \cite{stopyra2024towards} and \cite{junzhe2023accurate} where parameter inference is carried out at the density field level, thus requiring forward modelling to predict the field value, followed by a pixel-by-pixel comparison to determine the likelihood. Our approach does not demand a pixel-by-pixel comparison, since it operates within the latent space. This is a more generic approach compared to the field-level inference, but it needs  a more comprehensive study to understand how various parameters (such as minimum halo mass, FoF method), affect the results and potentially bias the constraints on $f\sigma_8$, therefore this represents a limitation for our current methodology.

For comparison, we also carried out an EFT approach and we found that the CNN performs better when inferring $f\sigma_8$. This out-performance may be attributed to the fact that the EFT is limited to wave numbers below $k < 0.2 \ \hmpc$, whereas the CNN extends its scale to values up to $k < 0.25\ \hmpc$. The broader range of wave numbers accessible to the CNN likely provides it with more information, enhancing its ability to infer $f\sigma_8$.

We leave for future work the extension of this analysis to models beyond the $\Lambda$CDM model, i.e. to eliminate the first assumption of current growth measurements as mentioned earlier since that requires N-body simulations for modified gravity. Also, we leave a more detailed comparison between the CNN and the conventional approach such as the Fisher matrix approach for future work, as it would require significant theoretical modifications on the non-linear part for the latter and more tests on the CNN architecture to investigate the effect on low redshifts. We also leave for future work the possible development of ML-based methods for parameter inference with real data.

Finally and to conclude, given that in a few years, the forthcoming galaxy surveys will provide a plethora of high-quality data related to the large-scale structure of the Universe, novel ways to analyze these data will be required to minimize the theoretical errors emanating from assumptions, just as the non-linear modelling at small scales. Here, we provided the first step in this direction, but more work will be required to bridge the gap between the theory and the actual data.

\begin{acknowledgments}
We acknowledge the i-LINK 2021 grant LINKA20416 by the Spanish National Research Council (CSIC), which initiated and facilitated this collaboration. We are also particularly indebted first and foremost to D.~Sapone, but also to C.~Baugh, C.~Carbonne, S.~Casas, J.~Garc\'ia-Bellido, K.~Ichiki, E.~Komatsu, E.~Majerotto, V.~Pettorino, A.~Pourtsidou, F.~Villaescusa - Navarro and T.~S.~Yamamoto for useful discussions. IO, SK and SN acknowledge support from the research project  PID2021-123012NB-C43 and the Spanish Research Agency (Agencia Estatal de Investigaci\'on) via the Grant IFT Centro de Excelencia Severo Ochoa No CEX2020-001007-S,
funded by MCIN/AEI/10.13039/501100011033. AJN acknowledges JSPS KAKENHI, Grant-in-Aid for Transformative Research Areas, 21H05454 and Grant-in-Aid for Scientific Research (C), 21K03625. SK is supported by the Spanish Atracci\'on de Talento contract no. 2019-T1/TIC-13177 granted by Comunidad de Madrid, the I+D grant PID2020-118159GA-C42 funded by MCIN/AEI/10.13039/501100011033, and Japan Society for the Promotion of Science (JSPS) KAKENHI Grant no. 20H01899, 20H05853, and 23H00110. IO is supported
by the fellowship LCF/BQ/DI22/11940033 from ”la Caixa” Foundation (ID 100010434) and by a Graduate Fellowship at Residencia de Estudiantes supported by Madrid City Council (Spain), 2022-2023.
KM would like to thank the “Nagoya University Interdisciplinary Frontier Fellowship” supported by Nagoya University and JST, the establishment of university fellowships towards the creation of science technology innovation, Grant Number JPMJFS2120.
Finally, the authors  acknowledge use of the Boltzmann code \texttt{CLASS}~\cite{Lesgourgues:2011re,Blas:2011rf} and the \texttt{python} packages \texttt{scipy}~\cite{2020SciPy-NMeth}, \texttt{numpy}~\cite{harris2020array} and \texttt{matplotlib}~\cite{Hunter:2007}.\\

\noindent {\it Numerical Analysis Files}: The numerical codes used in this analysis will be available upon publication at  \href{https://github.com/murakoya/fsigma8\_ML}{https://github.com/murakoya/fsigma8\_ML}\\
\end{acknowledgments}

\appendix

\section{Loss functions tests}
The MSE Loss calculates the mean squared error between the CNN estimates and the true values and is set up to find an approximation of the marginal posterior mean. In general, the marginal posterior describes the probability that a simulation (and its array of clustering measurements) were created with a particular combination of parameters $\theta_i$.

The advantage of LFI with respect to the other loss functions is that it achieves a better convergence between the CNN output and both the mean $\mu_i$ and standard deviation $\sigma_i$ of the marginal posterior parameters $\theta_{i,j}$. LFI eliminates the influence of the overall scale in the spread of a parameter by incorporating logarithms. This approach assigns weights based on inverse variance to the gradients of the various terms, in contrast to the MSE (Mean Squared Error) loss. Consequently, it optimizes the combination of gradients to account for each term's importance. According to the mentioned arguments, LFI Loss is defined by~\cite{perez2023constraining}: 

\begin{align}
\mathcal{L}_\mathrm{LFI}= &\log \left(\sum_{j \in \text { batch }}\left(\theta_{j}-\mu_{j}\right)^2\right) \nn \\
&+\log \left(\sum_{j \in \text { batch }}\left(\left(\theta_{j}-\mu_{j}\right)^2-\sigma_{j}^2\right)^2\right).
\end{align}

The input parameter of each training and test procedure was $f\sigma_8$. This particular loss function takes the logarithm of the squared difference between the true and predicted values of this parameter within a batch $j$ of the simulation, taking into account the standard deviation in the second term. As a matter of interest, we found a similar outcome with all the considered Loss functions - an error decrease for larger redshifts. It is worth mentioning that LFI performs better, as expected. In Table~\ref{LossFunctions}, we show the results only for the MSE and LFI Loss functions for simplicity, where we see that the LFI improves slightly the error estimates compared to the previously considered MSE Loss function.

\begin{table}[h]
  \centering
  \begin{tabular}{ c | c  c  c  c }
    \hline \hline
     redshift & MSE Loss & MAE Loss & Huber Loss & LFI Loss\\ \hline
    0.5 & 3.8 & 4.1 & 3.9 & 3.3 \\
    1.0 & 2.3 & 3.1 & 2.9 & 2.3 \\
    2.0 & 1.2 & 1.5 & 1.8 & 1.7 \\
    3.0 & 0.74 & 1.0 & 1.1 & 1.3 \\
    \hline \hline
  \end{tabular}
  
  \rightline{($\times 10^{-2}$)}
  \caption{Predicted errors for each redshift snapshot using Mean Squared Error (MSE), Mean Absolute Error (MAE), Hubber Loss and the moments networks Loss (LFI) Functions in the CNN architecture.}
  \label{LossFunctions}
\end{table}

It is worth mentioning that the slight improvement found at low redshift, achieved by the LFI loss, had a cost in stability, as mentioned before. This is why we chose the most stable method, MSE loss, for the final results.


\bibliography{main}

\end{document}